\documentclass[english]{article}
\usepackage[latin9]{inputenc}
\usepackage{geometry}
\usepackage{amsmath}
\usepackage{amssymb}

\usepackage{babel}

\makeatletter
\date{Feb 2010}

\makeatother

\begin{document}

\title{Permutation Models for Collaborative Ranking}

\author{Truyen Tran and Svetha Venkatesh\\
Department of Computing, Curtin University, Australia}
\maketitle
\begin{abstract}
We study the problem of collaborative filtering where ranking information
is available. Focusing on the core of the collaborative ranking process,
the user and their community, we propose new models for representation
of the underlying permutations and prediction of ranks. The first
approach is based on the assumption that the user makes successive
choice of items in a stage-wise manner. In particular, we extend the
Plackett-Luce model in two ways - introducing parameter factoring
to account for user-specific contribution, and modelling the latent
community in a generative setting. The second approach relies on log-linear
parameterisation, which relaxes the discrete-choice assumption, but
makes learning and inference much more involved. We propose MCMC-based
learning and inference methods and derive linear-time prediction algorithms. 
\end{abstract}

\section{Introduction}

Collaborative filtering is an important class of problems with the
promise to deliver personalised services. Members of communities rate
items in a service, and strong patterns exist between similar communities
of users. These patterns can be exploited to produce ranked lists
of items from a set of items not not previously exposed to the user.

Research in recommendation systems models user preferences through
a numerical rating - for example, rate a movie as $4$ or $5$ stars.
Although these users are forced into numeric scoring, these scores
are assigned qualitatively, and do not carry the assumed rigour of
quantitative evaluation. Also, this limits the expressiveness of preferences.
For example, a more intuitive way is to express the order of preferences
for a set of items. It may be easier to rank a set of movies, or the
top $10$ places visited, rather than assign them a numeric score.
Importantly, in recommendation systems, the core value proposition
is to recommend unseen items - this is where ranking rather than actual
rating becomes significant.

This paper addresses the open problem of recommending a ranked list
of items, or a preference list, without requiring intermediate ratings,
in collaborative filtering systems. Each user provides a ranked list
of items, in the decreasing order of preference. The list needs not
be complete, e.g. a user typically rates $10$ or $20$ items. The
intuition in collaborative filtering is that the community as a whole
may cover thousands of items, and as users belong to clusters within
this community, the properties of rankings within such clusters can
be transferred to a user for items that user has not seen. The technical
issue is to model the ranked item set both for a user and the community,
and predict the rank of unseen items for each user.

Despite of its importance, the collaborative ranking problem has only
been attempted recently \cite{weimer2008cofi,Weimer:2008xy,liu2009probabilistic}.
The papers \cite{Weimer:2008xy,liu2009probabilistic} consider pairwise
preferences , ignoring the simultaneous interaction between items.
Listwise approaches, studied in statistics (e.g. see \cite{mallows1957non,plackett1975analysis,diaconis1989generalization}),
often involve a relatively small set of items (e.g. in election, typically
less than a dozen of candidates are considered). Further, statisticians
are interested in the distribution of ranks in the population rather
than properties of individuals. Collaborative ranking, on the other
hand, differs in three ways: a) the scale is significantly different
- sometimes there are millions of items; b) the data is highly sparse,
that is a user will typically users will only express their preference
over a few items; and c) the personalisation aspect is crucial and
this the the distribution of rank per user is more important.

In this paper, focusing on the user, we study two approaches for modelling
the rank or preference lists. Our first approach assumes that the
user, when ranking items, will make successive choice in a stage-wise
matter. We extend one of the most well-known methods, namely the Plackett-Luce
model, to effectively model user-specific rank distribution in two
ways. First, we introduce parameter factoring into user-specific and
item-specific parameters. Second, we employ a generative framework
which models the community the user belongs to as a latent layer,
enabling richer modelling of the community structure in the ranking
generation process. We provide algorithms for learning the model parameters
and for ranking unseen items in linear time. The approach is detailed
in Section~\ref{sec:Latent-Discrete-Choice}.

The second approach relaxes the stage-wise choice assumption and models
intrinsic features of the permutation in a log-linear setting. Potentials
in the model capture the likelihood of an item in a specific position,
and for all item pairs, the likelihood of the first item being ordered
before the second. Although exact learning and inference is intractable,
we show that truncated MCMC techniques are effective for learning,
and for prediction that can be computed in linear time. The approach
is described in Section~\ref{sec:Log-linear-Models}.

The novelty in our contribution lies in the proposal of two approaches
incorporating key aspects of collaborative ranking: the user, their
specific communities, and the nature of the ranking list itself. The
work contributes efficient methods for learning and prediction.

\section{Preliminaries\label{sub:Pairwise-Losses}}

Suppose that we have a data set of $N$ users, and $M$ items and
each user $u\in\{1,2,...,N\}$ provides a list of $n_{u}\le M$ ranked
items $\pi^{u}=\{\pi_{1}^{u},\pi_{2}^{u},...\pi_{n_{u}}^{u}\}$, where
$\pi_{i}^{u}$ is the index of the item in position $i$. For notational
simplicity, we will drop the explicit superscript $u$ in $\pi^{u}$
when there is no confusion, and use $y=\pi_{i}$ when we mention the
item $y\in\{1,2,...,M\}$ in position $i$. The goal is to effectively
model the distribution $P(\pi|u)$. The main difficulty is that the
number of permutations is $n_{u}!$, which is only tractable for small
$n_{u}$.

A simplified way is to examine the ordering between only two items
(e.g. see \cite{Weimer:2008xy,liu2009probabilistic}). Denote by
$s_{\pi_{i}}^{u}$ the scoring function when the item is positioned
at $i$ in the list $\pi$ of user $u$. Let us consider the following
quantity 
\begin{eqnarray*}
d_{ij}^{u} & = & \mbox{sign}(j-i)(s_{\pi_{i}}^{u}-s_{\pi_{j}}^{u})\,.
\end{eqnarray*}
Basically $d_{ij}^{u}$ is positive when the scoring functions $\{s_{y}^{u},s_{y'}^{u}\}$
agree with their relative positions in the list, and negative otherwise.
For simplicity, we assume the factoring $s_{y}^{u}=\sum_{k=1}^{K}W_{uk}H_{ky}$
where $W\in\mathbb{R}^{N\times K}$ and $H\in\mathbb{R}^{K\times M}$
for some $K<\min\{M,N\}$. Thus the learning goal is to estimate $\{W,H\}$
so that $\{d_{ij}^{u}$\} are positive for all the triples $(u,i,j)$
in the training data, where $1\le i<j\le n_{u}$. This suggests a
regularised loss function in the form
\[
\mathcal{R}=\frac{1}{N}\sum_{u}\sum_{i=1}^{n_{u}}\sum_{j=i+1}^{n_{u}}L(d_{ij}^{u})+\Omega(W,H)\,,
\]
 where $L(d_{ij}^{u})$ is the user-specific loss and $\Omega(W,H)=\alpha\sum_{uk}W_{uk}^{2}+\beta\sum_{yk}H_{ky}^{2}$
is the regularising component. Popular choices of $L(d_{ij}^{u})$
are 
\[
L(d_{ij}^{u})=\begin{cases}
(1-d_{ij}^{u})^{2} & \,\,\,\mbox{in regression;}\\
\max(0,1-d_{ij}^{u}) & \,\,\,\mbox{in large-margin setting; and}\\
\log(1+\exp\{-d_{ij}^{u}\}) & \,\,\,\mbox{in logistic regression}.
\end{cases}
\]

\section{Latent Discrete Choice Models \label{sec:Latent-Discrete-Choice}}

We now address the listwise models, starting from the assumption that
the user makes the ranking decision in a stage-wise manner. We will
focus on the Plackett-Luce model \cite{plackett1975analysis}
\begin{equation}
P(\pi)=\prod_{i=1}^{M}\frac{e^{s_{\pi_{i}}}}{\sum_{j=i}^{M}e^{s_{\pi_{j}}}}\,,\label{eq:Plackett-Luce}
\end{equation}
where $s_{\pi_{i}}$ is the score associated with the item at position
$i$ in the permutation $\pi$. The probability that an item is chosen
as the first in the list is $e^{s_{\pi_{1}}}/\sum_{j=1}^{M}e^{s_{\pi_{j}}}$.
Once this item has been chosen, the probability that the next item
is chosen as the second in the remaining of $M-1$ item list is $e^{s_{\pi_{2}}}/\sum_{j=2}^{M}e^{s_{\pi_{j}}}$.
The process repeats until all items have been chosen in appropriate
positions.

However, this model is not suitable for collaborative ranking, because
it does not carry any personalised information, and lacks the concept
of community among users. We now introduce our extensions, first by
modelling the user-specific distribution $P(\pi|u)$ (Section~\ref{sub:Factored-Benter-Plackett-Luce-Model}),
and then proposing community-generated choice making (Section~\ref{sub:Latent-Semantic-Plackett-Luce}).

\subsection{Factored Benter-Plackett-Luce Model \label{sub:Factored-Benter-Plackett-Luce-Model}}

In collaborative ranking, we are interested in modelling the choices
by each user, and the permutation $\pi$ given by a user is incomplete
(i.e. the user often ranks a very small subset of items). We thus
introduce an user-specific model as

\begin{eqnarray*}
P(\pi|u) & = & \prod_{i=1}^{n_{u}}\frac{e^{s_{\pi_{i}}^{u}}}{\sum_{j=i}^{n_{u}}e^{s_{\pi_{j}}^{u}}}\,.
\end{eqnarray*}

Thus $s_{\pi_{i}}^{u}$ is the ranking score for item at position
$i$ (under $\pi$) by user $u$. However, this model does not account
for the the order at the beginning of the list being more important
than that at the end. We employ the technique by \cite{benter2008computer},
introducing \emph{damping} factors $\rho_{1}\ge\rho_{2}\ge...\ge\rho_{n}\ge0$
as follows

\[
P(\pi|u)=\prod_{i=1}^{n_{u}}\frac{e^{\rho_{i}s_{\pi_{i}}^{u}}}{\sum_{j=i}^{n}e^{\rho_{i}s_{\pi_{j}}^{u}}}\,.
\]
As an example, we may choose $\rho_{i}=1/\log(1+i)$.

In the standard Plackett-Luce model, the set of parameters $\{s_{y}\}$
can be estimated from a set of $i.i.d$ permutation samples. In our
adaptation, however, this trick does not work because the score $s_{y}^{u}$
will be undefined for unseen items. Instead, we propose to factor
$s_{y}^{u}$ as follows
\[
s_{y}^{u}=\sum_{k=1}^{K}W_{uk}H_{ky}\,,
\]
where $W\in\mathbb{R}^{N\times K}$ and $H\in\mathbb{R}^{K\times M}$
for some $K<\min\{M,N\}$ are parameter matrices. The $y$th column
of $H$ can be considered as the feature vector of item $y$, and
the $u$th row of $W$ as the parameter vector specific to user $u$.

To learn the model parameters, maximum likelihood estimation can be
carried out through maximising the following regularised log-likelihood
with respect to $\{W,H\}$
\[
\mathcal{L}(W,H)=\sum_{u}\log P(\pi|u)-\alpha\left\Vert W\right\Vert _{F}^{2}-\beta\left\Vert H\right\Vert _{F}^{2}\,,
\]
 for $\alpha,\beta>0$. It can be verified that the regularised log-likelihood
is concave in either $W$ or $H$, but not both. Once the model has
been specified, $\{s_{y}^{u}=\sum_{k=1}^{K}W_{uk}H_{ky}\}$ can be
used for sorting the items previously not seen by the user, where
larger $s_{y}^{u}$ ranks the item higher in the list.

\subsection{Latent Semantic Plackett-Luce Model \label{sub:Latent-Semantic-Plackett-Luce}}

The model in the previous subsection lacks generative interpretation-
we do not know how the ranking is generated by the user. A principled
way is to assume that the user belongs to hidden communities, and
that those communities will jointly generate the ranking. Recall that
in the Plackett-Luce model, the choice of items is made stage-wise
- the next item is chosen given that previously chosen items are ahead
in the list. Denote by $P_{i}(\pi|z,u)$ the probability of choosing
the item for the $i$th position by $u$ with respect to community
$z$, i.e.

\begin{equation}
P_{i}(\pi|z,u)=\frac{e^{s_{\pi_{i}}^{z}}}{\sum_{j=i}^{n_{u}}e^{s_{\pi_{j}}^{z}}}\,.\label{eq:Plackett-Luce-local}
\end{equation}

Let $P(z|u)$ be the probability that the user belongs to one of the
communities $z\in\{1,2,..,K\}$, then the user-specific permutation
is defined as

\begin{eqnarray}
P(\pi|u) & = & \prod_{i=1}^{n_{u}}\sum_{z}P(z|u)P_{i}(\pi|z,u)\,.\label{eq:Latent-Plackett-Luce}
\end{eqnarray}

Due to the sum in the denominator in Equation~\ref{eq:Plackett-Luce-local},
we may expect that the computation of $P(\pi|u)$ takes $n_{u}(n_{u}-1)K/2$
time. However, we can compute in $n_{u}K$ time by precomputing a
recursive array $A_{i}^{z}=A_{i+1}^{z}+e^{s_{\pi_{i}}^{z}}$ for $1\le i<n_{u}$.
If we start with $A_{n_{u}}=e^{s_{\pi_{n_{u}}}^{z}}$, then clearly
$A_{i}^{z}=\sum_{j=i}^{n_{u}}e^{s_{\pi_{j}}^{z}}$, which is the denominator
in Equation~\ref{eq:Plackett-Luce-local}.

\subsubsection{Learning using EM}

There are two sets of parameters to estimate, the mixture coefficients
$\{P(z|u)\}$ and the community-specific item scores $\{s_{y}^{z}\}$.
We describe an EM algorithm for learning these parameters, starting
from the lower-bound of the incomplete log-likelihood $\mathcal{L}=\sum_{u}\log P(\pi|u)$
as

\begin{eqnarray*}
\mathcal{L} & = & \sum_{u}\sum_{i=1}^{n_{u}}\log\sum_{z}P(z|u)P_{i}(\pi|z,u)\\
 & \ge & \sum_{u}\sum_{i=1}^{n_{u}}\sum_{z}Q_{i}(z|\pi,u)\log P(z|u)P_{i}(\pi|z,u)\\
 & = & \mathcal{Q}\,.
\end{eqnarray*}
 where $Q_{i}(z|\pi,u)$ is defined at each \textbf{E}-step $t+1$
as follows 
\[
Q_{i}^{t+1}(z|\pi,u)\leftarrow\frac{P^{t}(z|u)P_{i}^{t}(\pi|z,u)}{P_{i}^{t}(\pi|u)}\,.
\]

In the \textbf{M}-step, we fix $Q_{i}(z|\pi,u)$ and estimate $\{P(z|u),\, s_{y}^{z}\}$
by maximising $\mathcal{Q}$. We equip the lower-bound with the constraint
$\sum_{z}P(z|u)=1$ through the Lagrangian function $\mathcal{F}=\mathcal{Q}+\sum_{u}\mu_{u}(\sum_{z}P(z|u)-1)$
where $\{\mu_{u}\}$ are Lagrange multipliers. Setting the gradient
of the Lagrangian function
\begin{eqnarray*}
\frac{\partial\mathcal{F}}{\partial P(z|u)} & = & \sum_{i=1}^{n_{u}}Q_{i}(z|\pi,u)\frac{1}{P(z|u)}+\mu_{u}
\end{eqnarray*}
 to zeros and maintaining that $\sum_{z}P(z|u)=1$ would lead to
\begin{eqnarray*}
P(z|u) & \leftarrow & \frac{\sum_{i=1}^{n_{u}}Q_{i}(z|\pi,u)}{\sum_{z}\sum_{i=1}^{n_{u}}Q_{i}(z|\pi,u)}\\
 & = & \frac{1}{n_{u}}\sum_{i=1}^{n_{u}}Q_{i}(z|\pi,u)\,.
\end{eqnarray*}

This closed form update, however, does not apply to $\{s_{y}^{z}\}$.
Instead, we resort to the gradient-based method, where

\begin{eqnarray*}
\frac{\partial\mathcal{Q}}{\partial s_{y}^{z}} & = & \sum_{i=1}^{n_{u}}Q_{i}(z|\pi,u)\frac{\partial\log P_{i}(\pi|z,u)}{\partial s_{y}^{z}}\\
 & = & \sum_{i=1}^{n_{u}}Q_{i}(z|\pi,u)\{\delta_{\pi_{i}}^{y}-\frac{\sum_{j=i}^{n_{u}}e^{s_{\pi_{j}}^{z}}\delta_{\pi_{j}}^{y}}{\sum_{j=i}^{n_{u}}e^{s_{\pi_{j}}^{z}}}\}\,,
\end{eqnarray*}
 where $\delta_{\pi_{i}}^{y}=1$ if $y=\pi_{i}$ and $0$ otherwise.
Typically, we run only a few updates for $s_{y}^{z}$ per \textbf{M}-step.

\subsubsection{Prediction \label{sub:Latent-Plackett-Luce-Prediction}}

Given that models are fully specified, we want to output a ranked
list of unseen items for each user $u$. However, finding the optimal
ranking for an arbitrary set of items is generally intractable and
thus we resort to finding the rank of just one unseen item at a time,
given that the seen items have been sorted. In other words, we fix
the orders of the old items, and then introduce one new item into
the model, assuming that this introduction does not change the relative
orders of the old items. So the problem now reduces to finding the
position of the new item among the old items.

We repeat the process for all new items, and determine their positions
in the list. If the two new items are placed in the same position,
then their relative ranks will be determined by the likelihood of
their introductions.

Let $\pi'$ be the new list after introducing a new item. Denote by
$\pi_{i:j}$ the set of items whose positions are from $i$ to $j$
under $\pi$. Suppose that the new item is placed between the $(j-1)$th
and the $j$th items of the the old list $\pi$, and thus it is in
the $j$th position of the new list $\pi'$. Thus $\pi'_{1:j-1}=\pi_{1:j-1}$
and $\pi'_{j+1:n+1}=\pi_{j:n}$. We want to find
\[
j^{*}=\arg\max_{j}P(\pi'_{1:j-1},\pi'_{j},\pi'_{j+1:n+1}|u)\,,
\]
 where $P(\pi'_{1:j-1},\pi'_{j},\pi'_{j+1:n+1}|u)=$
\begin{eqnarray*}
\left[\prod_{i=1}^{j-1}\sum_{z}P(z|u)P_{i}(\pi'|z,u)\right]\left[\sum_{z}P(z|u)P_{j}(\pi'|z,u)\right] & \times\\
\times\left[\prod_{i=j+1}^{n+1}\sum_{z}P(z|u)P_{i}(\pi'|z,u)\right]\,.
\end{eqnarray*}

Naive computation for finding the optimal $j^{*}$ will cost $n_{u}(n_{u}+1)K/2$
steps. Here we provide a solution with just $(n_{u}+1)K$ steps. We
will proceed from left-to-right in a recursive manner, starting from
$j=1$. Recall that we can compute $P(\pi'_{1:n+1}|u)$ in Equation~\ref{eq:Latent-Plackett-Luce}
in $(n_{u}+1)K$ steps.

Assume that we have computed for the case that the position of the
new item is $j$ (under $\pi'$), we want to compute the case that
the new position is $j+1$ (under $\pi''$). Let us examine the odds
\[
O_{j}=\frac{P(\pi''_{1:j},\pi''_{j+1},\pi''_{j+2:n+1}|u)}{P(\pi'_{1:j-1},\pi'_{j},\pi'_{j+1:n+1}|u)}\,.
\]

We have $P(\pi''_{1:j},\pi''_{j+1},\pi''_{j+2:n+1}|u)=$ 
\begin{eqnarray*}
\left[\prod_{i=1}^{j-1}\sum_{z}P(z|u)P_{i}(\pi''|z,u)\right]\left[\sum_{z}P(z|u)P_{j}(\pi''|z,u)\right]\times\\
\times\left[\sum_{z}P(z|u)P_{j+1}(\pi''|z,u)\right]\left[\prod_{i=j+2}^{n+1}\sum_{z}P(z|u)P_{i}(\pi''|z,u)\right]\,.
\end{eqnarray*}
 We now notice that $\pi''_{1:j-1}=\pi'_{1:j-1}$ and $\pi''_{j+2:n+1}=\pi'_{j+2:n+1}$,
and 
\begin{eqnarray*}
P_{i}(\pi'|z) & = & P_{i}(\pi''|z)\\
\forall z,\,\,\mbox{and for}\,\, i & \in & \{1:j-1\}\cup\{j+2:n_{u}+1\}\,.
\end{eqnarray*}

The odds can be simplified as 
\begin{eqnarray}
O_{j}= & \frac{\left[\sum_{z}P(z|u)P_{j}(\pi''|z,u)\right]\left[\sum_{z}P(z|u)P_{j+1}(\pi''|z,u)\right]}{\left[\sum_{z}P(z|u)P_{j}(\pi'|z,u)\right]\left[\sum_{z}P(z|u)P_{j+1}(\pi'|z,u)\right]}\,.\label{eq:odd}
\end{eqnarray}
 which costs $K$ time to evaluate. Consequently, the recursive process
costs totally $(n_{u}+1)K$ time steps.

\section{Log-linear Models \label{sec:Log-linear-Models}}

In this section, we propose a second approach to permutation modelling.
The main difference from the first approach is that we do not make
the discrete-choice assumption, which makes the parameter estimation
easy, but complicates the inference. We now rely on the log-linear
parameterisation, which is more flexible. The generic conditional
distribution is defined as
\begin{equation}
P(\pi|u)=\frac{1}{Z(u)}\left[\prod_{i=1}^{n_{u}}\phi_{\pi}(i,u)\right]\left[\prod_{i=1}^{n_{u}-1}\prod_{j=i+1}^{n_{u}}\phi_{\pi}(i,j)\right]\,,\label{eq:log-linear}
\end{equation}
 where $\phi_{\pi}(i,u)$ and $\phi_{\pi}(i,j)$ are positive potential
functions, $Z(u)$ is the normalising constant (a.k.a partition function).
The position-wise potential $\phi_{\pi}(i,u)$ captures the likelihood
that a particular item $y=\pi_{i}$ is placed in position $i$ by
user $u$. For example, we would expect that a particular movie is
among the top $5\%$ in the list of a user. On the other hand, the
pairwise potential $\phi_{\pi}(i,j)$ encodes the likelihood that
the item $y=\pi_{i}$ is preferred to item $y'=\pi_{j}$. In what
follows, we will make use of the energy notion, i.e. $\phi_{\pi}(i,u)=\exp\{-E(\pi_{i},u)\}$
and $\phi_{\pi}(i,j)=\exp\{-E(\pi_{i},\pi_{j})\}$. The energy of
the permutation $\pi$ is therefore the sum of component energies,
i.e. $E(\pi,u)=\sum_{i}E(\pi_{i},u)+\sum_{i}\sum_{j>i}E(\pi_{i},\pi_{j})$.

\subsection{MCMC for Inference}

Inference in the above generic model is intractable due to the partition
function $Z(u)$, which requires $\frac{1}{2}n_{u}^{2}(n_{u}-1)^{2}(n_{u}-2)!$
computational steps%
\footnote{There are $n_{u}!$ permutations, each require $\frac{1}{2}n_{u}(n_{u}-1)$
steps of computing the product of potentials.%
}. We thus resort to MCMC methods. The key is to design a proposal
distribution that helps the random walks to quickly reach the high
density regions. There is also a trade-off here because large steps
would mean significant distortion of the current permutation, resulting
in more computational cost per move. We consider three types of local
moves.

\emph{Item relocation}. Randomly pick one item in the list, and relocate
it, keeping the relative orders of the rest unchanged. For example,
assume the permutation is $[A,B,C,D,E,F]$ and if $B$ is relocated
to the place between $E$ and $F$, then the new permutation is $[A,C,D,E,B,F]$.
Generally, this type of move costs $\mathcal{O}(n_{u})$ operations
per move due to the change in relative preference orders. In the example
we are considering, the pairs $BC,BD,BE$ would change to $CB,DB,EB$.

\emph{Item swapping}. Randomly pick two items, and swap their positions
leaving other items unchanged. In the above example, if we swap $B$
and $E$, then the new permutation is $[A,E,C,D,B,F]$. This also
costs $\mathcal{O}(n_{u})$ operations per move.

\emph{Sublist permutation}. Randomly pick a small sublist, try all
permutations within this sublist. For example, the sublist $[B,C,D]$
will result in $[C,B,D],[B,D,C],[D,C,B],[C,D,B],[D,B,C]$. This costs
$\Delta!$ where $\Delta$ is the size of the sublist. When $\Delta=2$,
this is the special case of the item swapping.

Since the proposals are symmetric, the acceptance probability in the
Metropolis-Hastings method is simply
\begin{equation}
P=\min\{1,e^{-\Delta E}\}\,,\label{eq:acceptance-rate}
\end{equation}
 where $\Delta E$ is the change in model energy due to the proposed
move.

\subsection{Learning with Truncated MCMC}

Learning using maximum likelihood is intractable due to the computation
of $Z(u)$ and its gradient, and thus MCMC-based learning can be employed.
The assumption is that if we generate enough samples according to
the model distribution, then the gradient of the log-likelihood can
be accurately estimated, and thus learning can proceed. However, this
is clearly too expensive, because generally we would need a significantly
large number of samples per gradient evaluation. Instead, Hinton \cite{Hinton02}
proposes a simple technique called Contrastive Divergence (CD) that
has been shown to work well in standard Boltzmann machines. The idea
is that instead of starting the Markov chain randomly and running
forever, we can just start from the observed configuration, and run
for a few steps. This is enough to relax the model from the empirical
distribution.

Here we adopt the CD, but we should stress in passing that the application
of CD in the context of permutation modelling is novel. It is possible
that we just need to run one short Markov chain of length $n_{u}$with
the item-swapping moves.

\subsection{Learning with Pseudo-likelihood}

In standard graphical models, pseudo-likelihood is an efficient alternative
to the full likelihood, and it is provably consistent given sufficient
regularity in the model structure. However, this concept has no straightforward
application in permutation models. We attempt to consider the pseudo-likelihood
concept from a more abstract level.

There is a close relationship between pseudo-likelihood and MCMC techniques.
The difference is that in MCMC we randomly choose one local permutation
configuration, while in pseudo-likelihood, we consider all local configurations,
and thus the process is deterministic. Using this idea, the (log)
pseudo-likelihood can be written as
\begin{eqnarray*}
\mathcal{L}^{pseudo} & = & \sum_{u}\sum_{c}\log P(\pi_{c}|\pi_{\neg c},u)\,\,\,\mbox{where}\\
P(\pi_{c}|\pi_{\neg c},u) & = & \frac{\exp\{-E(\pi_{c},\pi_{\neg c},u)\}}{\sum_{\pi'_{c}}\exp\{-E(\pi'_{c},\pi'_{\neg c},u)\}}\,.
\end{eqnarray*}
 and $c$ denotes the index of the local structure, and $\neg c$
denotes the rest of the items whose relative positions remain unchanged.
We briefly discuss three types of local structure.

\emph{Item relocation}. All the items will be considered, each has
the following local distribution

\[
P(\pi_{i}|\pi_{\neg i},u)=\frac{\exp\{-E(\pi_{1:i-1},\pi_{i},\pi_{i+1:n},u)\}}{\sum_{j=1}^{n}\exp\{-E(\pi'_{1:j-1},\pi'_{j},\pi'_{j+1:n},u)\}}
\]
 for $1\le i\le n_{u}$. Since the denominator is the sum over $n_{u}$
positions, each requires $n_{u}-1$ pairwise energies, naively computing
$P(\pi_{i}|\pi_{\neg i},u)$ would result in $n_{u}(n_{u}-1)$ steps.
However, we can the the denominator in a single pass. Suppose the
item $y=\pi_{i}$ moves from current position $j$ (under $\pi'$)
to $j+1$ (under $\pi''$), then change in energy is
\[
\Delta E_{j}(\pi'\rightarrow\pi'',u)=E(\pi''_{j},\pi''_{j+1},u)-E(\pi'_{j},\pi'_{j+1},u)\,,
\]
which costs a constant time to compute. We can start with $j=1$,
updating model energies in one pass.

\emph{Item swapping}. We have $n_{u}(n_{u}-1)/2$ item pairs for each
user $u$. So the local distribution is
\[
P(\pi_{i,j}|\pi_{\neg i,j},u)=\frac{1}{1+\exp\{-\Delta E_{ij}(u)\}}
\]
 for $1\le i<j\le n_{u}$ where $\Delta E_{ij}(u)$ is the change
in energy as a result of the swapping items $y=\pi_{i}$ and $y'=\pi_{j}$.

\emph{Sublist permutation.} We will have $n_{u}+1-\Delta$ local distributions
of the following form $P(\pi_{i:i+\Delta-1}|\pi_{\neg i:i+\Delta-1},u)=$
\[
\frac{\exp\{-E(\pi_{1:i-1},\pi_{i:i+\Delta-1},\pi_{i+\Delta:n},u)\}}{\sum_{\pi'_{j:j+\Delta-1}}\exp\{-E(\pi'_{1:j-1},\pi'_{j:j+\Delta-1},\pi'_{j+\Delta:n},u)\}}
\]
 for $1\le i\le n_{u}+1-\Delta$.

\subsection{Prediction}

We employ the same technique described earlier with the Latent Plackett-Luce
model (Section~\ref{sub:Latent-Plackett-Luce-Prediction}) in that
we fix the relative order of the items the user has already seen,
and introduce the new item into the list. Then we search for the best
position of the new item in the list, where the best position has
the lowest permutation energy. Computationally, this is similar to
the pseudo-likelihood with item-relocation, except that now we choose
the most probable position instead of summing over all positions.
Thus, we can find the best position in a single pass.

\subsection{Parameterisation Case Studies \label{sub:Parameterisation-Case-Studies}}

We now specify the parameters for the log-linear modelling. We will
focus on two special cases, one with factored position-wise parameters,
and the other with pairwise parameters.

\subsubsection{Factored Position-wise Parameters \label{sub:Factored-Position-wise-Parameters}}

Let us start from the idea of augmenting each item with a score $s_{y}^{u}$,
which we assume the factored form as $s_{y}^{u}=\sum_{k=1}^{K}W_{uk}H_{ky}$.
Ignoring the pairwise potentials in Equation~\ref{eq:log-linear},
the position-wise potential can be defined as $\phi_{\pi}(i,u)=\exp\{s_{\pi_{i}}^{u}g(i,u)\}$
where $g(i,u)$ is a monotonically \emph{decreasing} function in $i$.
This case is attractive because a MCMC step with position swapping
costs only a constant time, i.e. if we swap two items at positions
$l$ and $m$, the change in energy is $\Delta E_{lm}(u)=2(s_{\pi_{l}}^{u}-s_{\pi_{m}}^{u})(m-l)$.
In addition, prediction is rather simple as we just need to use $s_{y}^{u}$
for sorting.

In particular, we are interested in the case $g(i,u)=(1+n_{u}-2i)/n_{u}$
since it has a nice interpretation

\begin{eqnarray*}
P(\pi|u) & = & \frac{1}{Z(u)}\exp\{\frac{1}{n_{u}}\sum_{i=1}^{n_{u}}s_{\pi_{i}}^{u}(1+n_{u}-2i)\}\\
 & = & \frac{1}{Z(u)}\exp\{\frac{1}{n_{u}}\sum_{i=1}^{n_{u}-1}\sum_{j=i+1}^{n_{u}}(s_{\pi_{i}}^{u}-s_{\pi_{j}}^{u})\}\,,
\end{eqnarray*}
 which basically says that when $y=\pi_{i}$ is preferred to $y'=\pi_{j}$,
then we should have $s_{y}^{u}>s_{y'}^{u}$.

\subsubsection{Pairwise Parameters \label{sub:Pairwise-Parameters}}

We now consider the second special case, where the pairwise potential
is simply $\phi_{\pi}(i,j)=\exp\{\lambda_{yy'}\}$ subject to $y=\pi_{i}$
and $y'=\pi_{j}$. Note that $\lambda_{yy'}\ne\lambda_{y'y}$. Since
the total parameters can be as much as $M^{2}$, which is often too
large for robust estimation, we keep only the parameters of the item
pairs whose number of co-occurrences in the training data is larger
than a certain threshold. To account for missing pairs, we also use
the position-wise potential $\phi_{\pi}(i,u)=\exp\{\gamma_{\pi_{i}}g(i,u)\}$
with an extra parameter per item $\gamma_{y}$ (here $y=\pi_{i}$).
The distribution is now defined as

\[
P(\pi|u)=\frac{1}{Z(u)}\exp\left\{ \sum_{i=1}^{n_{u}}\gamma_{\pi_{i}}g(i,u)+\sum_{i=1}^{n_{u}-1}\sum_{j=i+1}^{n_{u}}\lambda_{\pi_{i},\pi_{j}}\right\} .
\]
For example, the threshold may be set to $5$ and we can use $g(i,u)=1-i/n_{u}$.
Note that there is no user-specific parameter. However, the distribution
is still user-dependent because of the number of items $n_{u}$ and
the ranking are user-specific.

In MCMC, suppose we swap items at positions $l$ and $m$, where $l<m$,
the change in energy is 
\begin{eqnarray*}
\Delta E_{lm}(u) & = & (\gamma_{\pi_{l}}-\gamma_{\pi_{m}})\{g(l,u)-g(m,u)\}+\\
 &  & \lambda_{lm}-\lambda_{ml}+\sum_{l<i<m}\{\lambda_{li}+\lambda_{im}-\lambda_{il}-\lambda_{mi}\}\,.
\end{eqnarray*}

\section{Related Work}

Although collaborative filtering with numerical ratings is well studied,
collaborative ranking is more recent. Work of \cite{weimer2008cofi}
introduces CoFi$^{RANK}$ - a non-probabilistic method which optimises
the bound of the NDCG score. The authors discuss several pairwise
loss functions mentioned in Section~\ref{sub:Pairwise-Losses}. An
adaptation of PLSA \cite{hofmann2001ulp} for pairwise preference
is given in \cite{liu2009probabilistic}. None of these papers attempt
to model the rank distributions.

In statistics, on the other hand, rank models are well-studied, some
of which we have already mentioned in previous sections: the Bradley-Terry
\cite{bradley1952rank} for pairwise preferences, the Plackett-Luce
\cite{plackett1975analysis} for discrete choices, the Mallows for
rank aggregation \cite{mallows1957non}, and the spectral decomposition
\cite{diaconis1989generalization}. Statistical data, however, is
often limited to small sets of items (e.g. less than a dozen) and
the goal is to model a single distribution for all users. Our work,
on the other hand, deals with large sets of sparsely ranked items,
and models user-specific distributions.

Rank learning has recently attracted much attention in Information
Retrieval. For example, the Plackett-Luce model has been adapted in
\cite{cao2007learning}. However, the setting is different, since
the items (e.g. documents or images) are associated with pre-computed
features, and the parameters are only associated with these features,
not the items. Collaborative ranking, on the other hand, discovers
these features directly from the data.

\section{Conclusion}

We have studied two approaches of permutation modelling for collaborative
ranking under different assumptions. The first approach follows the
Plackett-Luce's discrete-choice assumption. We introduce parameter
factoring as well as latent semantic extensions to account for hidden
community structure among users. The second approach relies on log-linear
parameterisation. We show how to perform MCMC-based inference, learning,
and efficient recommendation. Future directions include extensions
to deal with ties among ranks, and to incorporate correlation between
users.

\end{document}